\documentclass[twocolumn,showpacs,preprintnumbers,amsmath,amssymb,APSl,prd,nofootinbib,superscriptaddress]{revtex4-1}

\usepackage{dcolumn}
\usepackage{bm}
\usepackage{ifpdf}
\usepackage{hyperref}
\usepackage{dcolumn}
\usepackage{bm}
\usepackage[spanish,english]{babel}
\usepackage{amsfonts}
\usepackage{amssymb}
\usepackage{graphicx}
\usepackage[latin1]{inputenc}

\newcommand \be{\begin{equation}}
\newcommand \en{\end{equation}}
\newcommand \bea{\begin{eqnarray}}
\newcommand \ena{\end{eqnarray}}

\begin{document}

\title{Brane-world and loop cosmology from a gravity-matter coupling perspective.}

\author{Gonzalo J. Olmo} \email{gonzalo.olmo@csic.es}
\affiliation{Departamento de F\'{i}sica Te\'{o}rica and IFIC, Centro Mixto Universidad de
Valencia - CSIC. Universidad de Valencia, Burjassot-46100, Valencia, Spain}
\affiliation{Departamento de F\'isica, Universidade Federal da
Para\'\i ba, 58051-900 Jo\~ao Pessoa, Para\'\i ba, Brazil}
\author{D. Rubiera-Garcia} \email{drubiera@fisica.ufpb.br}
\affiliation{Departamento de F\'isica, Universidade Federal da
Para\'\i ba, 58051-900 Jo\~ao Pessoa, Para\'\i ba, Brazil}

\pacs{04.50.Kd, 04.70.Bw, 11.10.Lm}

\date{\today}

\begin{abstract}
We show that the effective brane-world and the loop quantum cosmology background expansion histories can be reproduced from a modified gravity perspective in terms of an $f(R)$ gravity action plus a $g(R)$ term non-minimally coupled with the matter Lagrangian. The reconstruction algorithm that we provide depends on a free function of the matter density that must be specified in each case and allows to obtain analytical solutions always.  In the simplest cases, the function $f(R)$ is quadratic in the Ricci scalar, $R$, whereas $g(R)$ is linear. Our approach is compared with recent results in the literature. We show that working in the Palatini formalism there is no need to impose any constraint that keeps the equations second-order, which is a key requirement for the successful implementation of the reconstruction algorithm.
\end{abstract}

\maketitle

\section{Introduction}

In the context of the early-time cosmology, the approaches based on brane-world models \cite{Shiromizu:1999wj} and loop quantum gravity \cite{Singh06} seem to favour a specific model of cosmic evolution characterized by quadratic relations between the Hubble parameter and the energy density of the fluid, dubbed \emph{quadratic cosmology}. In a recent paper, Bertolami and P\'aramos \cite{Bertolami} considered theories in which, besides an $f(R)$ action for the gravitational field, the matter was also allowed to couple non-minimally to gravity via another $g(R)$ function. They showed that these theories may successfully reproduce any background expansion history. In particular, they found specific forms for the functions $f(R)$ and $g(R)$ able to reproduce quadratic cosmology in a four-dimensional scenario. An interesting aspect of the approach by Bertolami and  P\'aramos is the fact that the expression for the functions $f(R)$ and $g(R)$ can be found analytically and take the simple forms $f(R)=R+\alpha R^2$ and $g(R)=1+\beta R$, with $\alpha$ and $\beta$ being specific constant parameters. The fact that a quadratic $f(R)$ Lagrangian can be directly related with the cosmology of loop quantum cosmology and braneworlds through the addition of a non-trivial coupling between matter and curvature is remarkable.

The approach followed in \cite{Bertolami} to obtain the functions $f(R)$ and $g(R)$ is, however, not fully satisfactory. In fact, as is well known, $f(R)$ theories (with or without matter-curvature couplings) are generically governed by fourth-order field equations, which makes it extremely difficult to find exact solutions. The higher-order derivatives can also be interpreted as representing new dynamical degrees of freedom associated to a scalar field. This scalar field is a function of the Ricci scalar in the case of pure $f(R)$ gravity but also involves the matter Lagrangian in the non-minimally coupled case at hand. In order to avoid the difficulties derived from the existence of higher-order derivatives or new scalar degrees of freedom, the strategy proposed in \cite{Bertolami} was to introduce a specific constraint between the functions defining the model in order to remove the new degrees of freedom (or, equivalently, the higher-order derivatives) from the field equations. In our view, this procedure is removing in an {\it ad hoc} manner a basic feature and key defining aspect of the theory. In fact, it represents an unnecessary act of violence aimed at forcing a theory to do something that in natural conditions it would not do. Moreover, given that the theory contains new dynamical degrees of freedom by construction, one might expect that small perturbations could excite those degrees of freedom thus making unstable the choice proposed in \cite{Bertolami}. Though for some particular models it could be robust, there is no guarantee that this strategy could be valid in general.

In this work, we show that a natural alternative formulation of the problem presented above exists. The key difference between our approach and that of Bertolami and P\'{a}ramos \cite{Bertolami} is that we work in the Palatini formalism, i.e., we assume that metric and connection are {\it a priori} independent geometrical objects \cite{Olmo:2012yv}. We find that working in the Palatini formalism, the restriction imposed in \cite{Bertolami} is not essential, as the field equations that one obtains are of second-order by construction. We note that the Palatini formulation was also a key element in \cite{Olmo:2008nf} to obtain an effective action of the $f(R)$ type able to capture the full dynamics of loop quantum cosmology, from the GR limit at low energies to the nonperturbative regime at the bounce. In that approach, curvature-matter couplings were not considered. Here we allow for this possibility and explore its effects and implications.

In the approach of \cite{Bertolami}, the geometry is implicitly assumed to be Riemannian, i.e.,  the connection is constrained \emph{a priori} to be given by the Christoffel symbols of the metric, which is the origin of the higher-order field equations arising in most theories of extended gravity. However, on geometrical grounds, metric and connection are equally fundamental and independent entities, carrying very different geometrical meanings. In this sense, the question on whether the underlying geometry of space-time is Riemannian or otherwise is not a matter of conventions but a foundational issue of gravitational physics that must be answered by experiments. In the case of classical GR, this question is irrelevant because if the connection is taken to be independent, its variation leads to a new equation whose solution is the Levi-Civita connection, thus yielding the same field equations as in the standard (metric) approach. This result follows from the particular functional form of the Einstein-Hilbert Lagrangian but, in general, it does not hold when one moves away from GR (with the main exception been Lovelock theories \cite{Borunda}). When $f(R)$ extensions are considered in the Palatini approach, the connection satisfies a set of algebraic (not differential) equations. The solution can be expressed as the Levi-Civita connection of an auxiliary metric $h_{\mu\nu}$ conformally related with the space-time metric $g_{\mu\nu}$. When non-minimal coupling is allowed, the conformal factor is a function that depends on the Lagrangian $f(R)$, the function $g(R)$, and the matter sources. The resulting equations for the metric are second order, like in the minimally coupled case with $g(R)=1$. As a result, the solution to the problem of finding a pair of functions $f(R)$ and $g(R)$ able to reproduce a particular cosmological background history does not require imposing the constraint proposed in \cite{Bertolami}.

There are several reasons that motivate the study of the matter-curvature coupling within the Palatini framework (see \cite{Bertolami4} for a discussion on these theories within the metric approach). It is well-known from the first recent studies of $f(R)$ theories \cite{Olmo:2005hc} (see also the review \cite{Olmo:2011uz}), that the gravitational field in Palatini theories depends intimately on the local energy-momentum {\it density} distributions, i.e., it is not just determined by the total amounts of energy-momentum in a given region \cite{Olmo:2006zu,Olmo:2008ye}. The details of how that energy-momentum is distributed does have an impact on the metric locally.
In fact, in simple models of black hole formation, it has been shown by means of exact analytical solutions \cite{MartinezAsencio:2012xn, Lobo:2013vga} (see also  \cite{Olmo:2011sw} for a perturbative discussion) that the space-time metric not only depends on the total mass of the collapsing fluid but also on the energy density that the fluid carries at each instant of time. When the energy flux that forms the black hole ceases, the resulting geometry only depends on the total accumulated mass. When the flux is on, a dependence on the energy-density appears again along the fluid's trajectory.
This puts forward that non-trivial interactions between geometry and energy-density arise even if one assumes minimal coupling. Furthermore, from the study of scenarios involving the coupling of high-energy extensions of gravity to free electric fields \cite{or}, it has been found that point-like particles could be seen as topological entities with wormhole structure \cite{Lobo:2014fma}. This non-trivial interplay between matter fields and geometry in which particles are seen as microscopic geometric structures \cite{Wheeler} naturally motivates the study of curvature-matter couplings as a way to encode high-energy interactions between geometry and topology.  Further research to better understand the role and properties of theories with this type of couplings is thus necessary.

In this work we consider a modified $f(R)$ gravity coupled to matter via a function $g(R)$ in the Palatini formalism. We shall explicitly show that the field equations of this setting are always second-order and, in vacuum, boil down to those of GR. This allows to study the gravity-matter coupling framework from a more general perspective than in the standard (metric) approach, since the functions $f(R)$ and $g(R)$ are not forced to satisfy a specific constraint.
Here we shall work out this scenario and show that any given cosmological background history can be obtained from a Palatini $f(R)$ theory with gravity-matter couplings. To illustrate this point we will consider the particular case of quadratic cosmology.

\section{Palatini theories with gravity-matter coupling}

The action defining our theory is written as follows

\be \label{eq:action}
S= \int d^4x \sqrt{-g} \left[\frac{f(R)}{2\kappa^2} + g(R) L_m (\psi_m,g_{\mu\nu}) \right],
\en
where $\kappa^2$ is a constant with suitable dimensions (in GR, $\kappa^2  \equiv 8\pi G/c^3$), $\sqrt{-g}$ is the determinant of the space-time metric $g_{\mu\nu}$, $f(R)$ and $g(R)$ are two arbitrary functions of the Ricci scalar $R=g^{\mu\nu}R_{\mu\nu}(\Gamma)$, constructed with the independent connection $\Gamma \equiv \Gamma_{\mu\nu}^{\lambda}$, and $L_m$ is the matter Lagrangian, where $\psi_m$ denotes collectively the matter fields, which are only coupled to the metric for simplicity. To obtain the field equations for the action (\ref{eq:action}) we perform independent variations with respect to metric and connection (Palatini approach), and further assume vanishing torsion, $\Gamma_{[\mu\nu]}^{\lambda}=0$ \cite{Torsion}, which leads to

\bea
(f_R+2\kappa^2 g_R L_m)R_{\mu\nu}-\frac{1}{2} g_{\mu\nu}f(R)&=&\kappa^2 g(R) T_{\mu\nu} \label{eq:metric}\\
\nabla_{\mu}^{\Gamma} \left[ \sqrt{-g} \left(f_R + 2\kappa^2 g_R L_m \right)g^{\alpha \beta} \right]&=&0, \label{eq:connection}
\ena
where we have used the short hand notation $f_R \equiv df/dR$ and $T_{\mu\nu}=-\frac{2}{\sqrt{-g}} \frac{\delta(\sqrt{-g}L_m)}{\delta g^{\mu\nu}}$ is the energy-momentum tensor of the matter. To solve these equations we note that the connection equations (\ref{eq:connection}) can be solved by introducing a rank-two tensor $h_{\mu\nu}$, related to the metric $g_{\mu\nu}$ as

\be \label{eq:sigma}
h_{\mu\nu}= \Phi g_{\mu \nu} \hspace{0.1cm} ; \hspace{0.1cm} h^{\mu\nu}= \frac{1}{\Phi} g^{\mu \nu},
\en
where

\be\label{eq:Phi}
\Phi \equiv f_R +2\kappa^2 g_R L_m,
\en
such that (\ref{eq:connection}) reads $\nabla_{\mu}^{\Gamma} \left(\sqrt{-h} h^{\alpha \beta} \right)=0$. This implies that the independent connection, $\Gamma_{\mu\nu}^{\lambda}$, becomes the Levi-Civita connection of $h_{\mu\nu}$, which is conformally related to the metric $g_{\mu\nu}$ via (\ref{eq:sigma}). Note that the GR case with no matter-curvature coupling corresponds to $f(R)=R$ and $g(R)=1$, which implies that $\Phi=1$ and therefore $g_{\mu\nu}=h_{\mu\nu}$, in agreement with the fact that in this case the action (\ref{eq:action}) is the standard Einstein-Hilbert action of GR.

On the other hand, tracing with $g_{\mu\nu}$ in (\ref{eq:metric}) we obtain
\be \label{eq:trace}
\Phi R-2f(R)=\kappa^2 g(R) T,
\en
where $T \equiv {T_\mu}^{\mu}$ is the trace of the energy-momentum tensor. This is an algebraic equation that generalizes the linear relation $R=-\kappa^2 T$ found in GR, and establishes a nonlinear relation between $R$ and $T$ and $L_m$ that depends on the form of both $f(R)$ and $g(R)$.

Raising an index in the metric equation (\ref{eq:metric}) with $h^{\mu\nu}$, it is easily seen that this equation reduces to
\be \label{eq:Rmunu}
{R_\mu}^{\nu}(h)=\frac{\kappa^2}{\Phi^2} \left[\frac{f(R)}{2\kappa^2} \delta_{\mu}^{\nu} +g(R){T_{\mu}}^{\nu} \right] \ ,
\en
which shows that $h_{\mu\nu}$ satisfies a set of second-order Einstein-like field equations. Given the fact that $h_{\mu\nu}$ is algebraically (conformally) related to $g_{\mu\nu}$ via the matter sources, the field equations will be second-order for $g_{\mu\nu}$ as well. From the trace equation (\ref{eq:trace}) it is also easily seen that in vacuum, ${T_\mu}^{\nu}=0$, one must have $R=R_0=$constant. As a consequence, the right-hand side of (\ref{eq:Rmunu}) turns into an effective cosmological constant term. This implies that the  equations boil down to those of GR with a cosmological constant term if $f(R_0)\neq 0$. Consequently no extra propagating degrees of freedom appear and the theory is free of ghost-like instabilities.

For operational purposes, it is also convenient to write the set of equations (\ref{eq:Rmunu}) in terms of ${G_\mu}^{\nu}(h)={R_\mu}^{\nu}(h)-\frac{1}{2} {\delta_\mu}^{\nu}R(h)$ as
\be \label{eq:Gmunu}
{G_\mu}^{\nu}(h)=\frac{\kappa^2}{\Phi^2} \left[g(R)\left({T_{\mu}}^{\nu}-\frac{1}{2}{\delta_\mu}^{\nu}T \right)-\frac{f(R)}{2\kappa^2} {\delta_\mu}^{\nu}  \right],
\en
which will be useful for some calculations in the following sections.

\section{Cosmic background evolution}

We now consider the case of a perfect fluid, with energy-momentum tensor
\be \label{eq:fluid}
{T_\mu}^{\nu}=(\rho +p)u_{\mu} u^{\nu} +p\delta_{\mu}^{\nu},
\en
where the unit vector $u_{\mu}$ satisfies $u_{\mu}u^{\mu}=-1$, while $\rho$ and $p$ are the energy density and pressure of the fluid, respectively. We are interested in a Friedmann-Robertson-Walker (FRW) scenario, corresponding to a homogeneous and isotropic universe, with line element for $g_{\mu\nu}$ given by
\be \label{eq:frw}
ds^2=g_{\mu\nu}dx^{\mu}dx^{\nu}=-dt^2+a^2(t)\delta_{ij} dx^i dx^j,
\en
Using the relation (\ref{eq:sigma}), we find that the components of $h_{\mu\nu}$ are
\be \label{eq:h}
h_{tt}=-\Phi \hspace{0.1cm}; \hspace{0.1cm} h_{ij}=\Phi a^2 \delta_{ij}.
\en
Using (\ref{eq:h}) we can easily calculate $G_{tt}(h) = 3(H+\frac{\dot{\Phi}}{2\Phi})^2$ (where $H=\dot{a}/a$ is Hubble's constant and a dot means a derivative with respect to $t$). Raising an index with $h^{tt}$, replacing the resulting expression into the left-hand-side of the field equations (\ref{eq:Gmunu}), and that of the energy-momentum tensor of the fluid (\ref{eq:fluid}) into the right-hand-side yields
\be \label{eq:Hubble1}
\left(H+\frac{\dot{\Phi}}{2\Phi} \right)^2=\frac{\kappa^2}{6\Phi} \left[g(\rho+3p)+\frac{f}{\kappa^2} \right] \ .
\en
Using Eq.(\ref{eq:trace}), we can remove the explicit dependence on the pressure from this equation to get
\be \label{eq:Hubble}
\left(H+\frac{\dot{\Phi}}{2\Phi}\right)^2=\frac{1}{6\Phi} \left[2 \kappa^2 g\rho -f+R\Phi \right] \ .
\en
This equation must be supplemented with the conservation equation, which is obtained from the Bianchi identities, $\nabla_{\mu}^h {G^\mu}_{\nu}=\nabla_{\mu}^g {G^\mu}_{\nu}+C^{\mu}_{\mu\lambda}{G^\lambda}_{\mu}-C^{\lambda}_{\mu \nu}{G^\mu}_{\lambda}=0$ ($\nabla_{\mu}^h$ and $\nabla_{\mu}^g$ being the covariant derivative with respect to the metric $h_{\mu\nu}$ and $g_{\mu\nu}$, respectively), where the last equality follows from the fact that $\Gamma_{\mu\nu}^{\lambda}$ is the Levi-Civita connection of $h_{\mu\nu}$. The tensor $C^{\lambda}_{\mu\alpha}=\Gamma_{\mu\alpha}^{\lambda}-L_{\mu\alpha}^{\lambda}$, where $L_{\mu\alpha}^{\lambda}$ is the Levi-Civita connection of $g_{\mu\nu}$, takes the form  $C_{\mu\alpha}^{\lambda}=\frac{1}{2}(\Phi_{\mu}\delta_{\alpha}^{\lambda} + \Phi_{\alpha}\delta_{\mu}^{\lambda} - \Phi^{\lambda}g_{\mu\alpha})$, where we have defined $\Phi^{\lambda} \equiv g^{\lambda \rho}\partial_{\rho} \log \Phi$. After some lengthy but straightforward algebra where we use both the field equations (\ref{eq:Gmunu}) and the relation between the metrics $g_{\mu\nu}$ and $h_{\mu\nu}$ in (\ref{eq:sigma}) we arrive at
\be \label{eq:conservation}
\nabla_{\alpha} {T^\alpha}_{\nu}=-(\partial_{\alpha} \log g(R)) [{T^\alpha}_{\nu}-\delta^{\alpha}_{\nu} L_m] \ ,
\en
which is in agreement with the results of \cite{Harko}. To fully specify this equation we need the explicit expression of the matter Lagrangian $L_m$. At this point there is a degeneracy, a problem already present in GR, as two formulations for the Lagrangian density of a perfect fluid are possible \cite{Hawking}:

\be
L_m=-\alpha \rho,
\en
where either $\alpha=1$ or $\alpha=-\omega$, with $\omega=p/\rho$ being the equation of state of the fluid.

From the above discussion, using the components of ${T_\mu}^{\nu}$ in (\ref{eq:fluid}), the $\nu=t$ component of the conservation equation (\ref{eq:conservation}) leads to

\be \label{eq:conservation}
\dot{\rho} \left[1+(1-\alpha)\frac{pg_RR_{\rho}}{g} \right]=-3H(\rho+p).
\en
It has been argued in \cite{Bertolami2}, on the basis of gravity-coupling models where non-geodesical motion and extra force may arise, that the physical sensible choice for this kind of problem corresponds to $\alpha=1$. Therefore we shall stick ourselves to that choice from now on. This implies $\dot{\rho}=-3H(1+\omega)\rho$ in the above equation, which coincides with the usual conservation law of the minimally coupled case. On the other hand, since for a perfect fluid with $p/\rho=\omega=$ constant the trace equation (\ref{eq:trace}) implies that $R=R(\rho)$, we can assume that $\Phi=\Phi(\rho)$ and thus $\dot{\Phi}=\Phi_{\rho} \dot{\rho}$. Replacing this into (\ref{eq:Hubble}) we obtain
\be \label{eq:Hsquare}
H^2= \frac{f(R)+ \kappa^2 g(R) (1+3\omega)\rho}{6\Phi(1-\Delta)^2},
\en
where we have defined $\Delta \equiv \frac{3}{2} (1+\omega)\rho \frac{\Phi_{\rho}}{\Phi}$ for notational simplicity. Note that when $f(R)=R$ and $g(R)=1$ the GR expression, $H^2_{GR}=\kappa^2 \rho/3$, is recovered.

Using results from \cite{Olmo:2011uz} we can compute the components $R_{tt}(h)$ and $R_{ij}(h)$ appearing on the left-hand side of the field equations (\ref{eq:Rmunu}). Equating to the right-hand side for the matter-energy source (\ref{eq:fluid})  yields
\bea
\dot{H}&+&\frac{\ddot{\Phi}}{2\Phi}+H^2+\frac{H\dot{\Phi}}{2\Phi} -\frac{1}{2} \left(\frac{\dot{\Phi}}{\Phi} \right)^2 \nonumber \\
&=& -\frac{1}{6\Phi} \left[2\kappa^2 g(R)\rho - f(R) \right] \\
\dot{H}&+&3H^2+\frac{\ddot{\Phi}}{2\Phi}+\frac{5H\dot{\Phi}}{2\Phi}  \nonumber \\
&=& \frac{1}{2\Phi} \left[2\kappa^2 g(R)p+f(R) \right] \ .
\ena
The subtraction of the first of these equations from the second leads to Eq.(\ref{eq:Hubble}), as can be easily verified. On the other hand, if we add five times the second equation to the first one we are led to
\bea \label{Hdot}
\dot{H}(1-\Delta)&+&\frac{H^2}{2} [1-\Delta^2 +6(\rho+p) \Delta_{\rho} ] \nonumber \\
&=&\frac{1}{12 \Phi}[f(R)-\kappa^2 g(R)(5\rho+3p)]
\ena
where one can show that $\Delta_{\rho}=\frac{\Delta}{\rho} +\frac{3}{2} (1+\omega)\rho (\frac{\Phi_{\rho\rho}}{\Phi} -(\frac{\Phi_{\rho}}{\Phi})^2) $. Once the functions $f(R)$ and $g(R)$ are specified, Eqs.(\ref{eq:Hsquare}) and (\ref{Hdot}) completely describe the cosmic expansion. One may also run the field equations in the opposite way, namely, propose a cosmological model as given by a certain function $H(\rho)$ and obtain the $f(R)$ and $g(R)$ functions that generate it. In the next section we shall consider this reconstruction scheme, illustrated with the application to the particular case of quadratic cosmology.

\section{Reconstructing quadratic cosmology}

In this section our model will be that of quadratic cosmology, defined by the functions
\bea
H^2&=& \frac{\kappa^2}{6} \rho \left(1+\epsilon \frac{\rho}{\rho_c} \right) \label{eq:H2a} \\
\dot{H}&=& -\frac{\kappa^2}{4}(1+\omega)\rho \left(1+2\epsilon \frac{\rho}{\rho_c} \label{eq:Hda} \right),
\ena
where $\epsilon= \pm 1$, $\omega=p/\rho$ is the equation of state and $\rho_c$ is some density scale where the corrections with respect to the standard GR prediction begin to play a role. Note that the expression for $\dot H$ in (\ref{eq:Hda}) comes from taking a derivative with respect to $t$ in  (\ref{eq:H2a}) and using the conservation equation. The interest in this model  stems from the fact that the quadratic density corrections to the GR dynamics with $\epsilon=+1$ arise in the context of brane-world scenarios \cite{Shiromizu:1999wj}, whereas the case with $\epsilon=-1$ is found within loop quantum gravity \cite{Singh06}. This puts forward that quadratic density corrections are very fundamental, as they appear in very different scenarios aimed at capturing new gravitational physics at high energies.

Our aim now is to show that the background cosmic dynamics of quadratic cosmology can be reproduced by Palatini $f(R)$ theories with a non-minimal gravity-matter coupling of the form discussed above. To proceed, we will replace the form of $H^2(\rho)$ and  $\dot H(\rho)$ given in (\ref{eq:H2a})  and (\ref{eq:Hda}), respectively, into (\ref{eq:Hsquare}) and (\ref{Hdot}). Since for a given equation of state $\omega$ Eq.  (\ref{eq:trace}) establishes an algebraic relation between $\rho$ and $R$, one can find expressions for $\rho$ and $R_\rho\equiv dR/d\rho$ as functions of $R$, $g$, $f$, and their derivatives with respect to $R$. Using these results in  (\ref{eq:Hsquare}) and (\ref{Hdot}) one could expect to find two equations that allow to solve for $f(R)$ and $g(R)$ (by means of numerical methods, at least). This approach would be similar to that used in \cite{Olmo:2008nf} to obtain the effective action of loop quantum cosmology, though in that case only the function $f(R)$ was necessary. The problem here is that the derivatives of $f(R)$ and $g(R)$ always appear through the combination (\ref{eq:Phi}) and, therefore, one cannot get independent equations for $f_{RRR}$ and $g_{RRR}$, which are the highest-order derivatives that appear in  the $\Phi_{\rho \rho}$ term of (\ref{Hdot}). Therefore, a different strategy is necessary.

The impossibility of getting independent equations for $f(R)$ and $g(R)$ stems from the fact that  (\ref{eq:trace}) is already establishing an algebraic relation between these two functions and their first derivatives with respect to $R$. Bertolami and P\'{a}ramos found that by imposing the condition $\Phi=$ constant the problem of finding $f(R[\rho])$ and $g(R[\rho])$ could be solved. In their approach, this choice has two effects. On the one hand, it avoids higher-order derivatives in the equations of motion, which would make the problem much more difficult, because all the higher-order derivatives appear acting on the function $\Phi$. On the other, it reduces the problem to finding just one function, as one of them can be eliminated from the equations using the constraint $\Phi=$ constant. In our case, we do not need to get rid of higher-order derivatives simply because they are not present in our problem from the very beginning. However, establishing a constraint between $f(R)$ and $g(R)$ through the function $\Phi$ does appear as a clever choice to simplify the analysis. We will thus assume in what follows that $\Phi=\Phi(\rho)$ is some given function of $\rho$.

The next step requires finding an expression for the Ricci scalar $R$ that appears in (\ref{eq:trace}) in terms of known quantities. Since $R=g^{\mu\nu}R_{\mu\nu}(\Gamma)$ and $R_{\mu\nu}(\Gamma)=R_{\mu\nu}(h)$, we can use the relation between the Ricci tensors of two conformally related metrics (see, for instance, appendix D in \cite{wald84}) to get
\be
R=R(g)+\frac{3}{2} \left( \frac{\dot{\Phi}}{\Phi} \right)^2 +\frac{3}{\Phi} (\ddot{\Phi} + 3H\dot{\Phi}) \ ,
\en
where $R(g)=6[\dot{H} + 2H^2]$. Since $\Phi=\Phi(\rho)$, we can use
  $\dot{\rho}=-3H(1+\omega)\rho$ to write $\dot \Phi=\Phi_\rho \dot\rho$, which leads to
\bea \label{eq:R}
R&=&6\dot{H}(1-\Delta)+12H^2 \Big[1+\frac{9}{4} \omega(1+\omega) \frac{\rho \Phi_{\rho}}{\Phi} \nonumber \\&+&
\frac{9}{4} (1+\omega)^2 \rho^2 \Big(\frac{\Phi_{\rho\rho}}{\Phi}+\frac{1}{2} \Big(\frac{\Phi_{\rho}}{\Phi}\Big)^2 \Big) \Big].
\ena
From this equation, using the form of $H^2(\rho)$ and  $\dot H(\rho)$ given in (\ref{eq:H2a})  and (\ref{eq:Hda}), respectively, and specifying a function $\Phi=\Phi(\rho)$ we find $R=R(\omega,\rho)$.

We now use the trace equation (\ref{eq:trace}) to isolate $f$ as
\be \label{eq:ff}
f(\rho)=\frac{R\Phi +\kappa^2 g(\rho) (1-3\omega)\rho}{2}.
\en
Replacing this equation into (\ref{eq:Hsquare}) we obtain
\be \label{eq:g}
g(\rho)=\Phi \frac{12H^2(1-\Delta)^2 -R}{3\kappa^2(1+\omega)\rho}.
\en
Putting this result back into (\ref{eq:ff}) we get
\be \label{eq:f}
f(\rho)=\frac{\Phi}{3(1+\omega)} [6H^2(1-\Delta)^2(1-3\omega)+R(1+3\omega)].
\en
Eqs.(\ref{eq:f}) and (\ref{eq:g}), together with the expression of the curvature scalar $R$ in (\ref{eq:R}) provide a full solution once a function $\Phi(\rho)$ is specified.
A representation of $f(R)$ and $g(R)$ is then possible, at least in parametric form. Let us now consider some illustrative examples.

\subsection{$\Phi=$constant} \label{phi1}

 As pointed out above, the choice $\Phi=$constant cancels out the fourth-order terms in the field equations of the metric approach studied in \cite{Bertolami}. As a result, the equations become identical with those found here in the Palatini formalism. As a consequence, the solution of \cite{Bertolami} for $\Phi=a=$ constant must be  equivalent to the one obtained here working in the Palatini approach. This can be easily checked by noting that in this case we have $\Delta=0$, which from formulae (\ref{eq:R}), (\ref{eq:f}) and (\ref{eq:g}),  and with the quadratic gravity functions (\ref{eq:H2a}) and (\ref{eq:Hda}) leads to
\bea
R&=&\frac{\kappa^2 \rho}{2} \left[ (1-3\omega) -2(1+3\omega) \epsilon \frac{\rho}{\rho_0} \right] \label{b1} \\
f(\rho)&=&\frac{\kappa^2 a \rho}{2} \left[1+3\omega \left(1+2\epsilon \frac{\rho}{\rho_0} \right) \right] \label{b2}  \\
g(\rho)&=&\frac{a}{2} \left(1+2\epsilon \frac{\rho}{\rho_0} \right), \label{b3}
\ena
From these equations the functions $f(R)$ and $g(R)$ can be directly obtained, i.e., by solving (\ref{b1}) to obtain $\rho=\rho(R)$ and then replace it into the expressions of $f(\rho(R))$ and $g(\rho(R))$, or parametrically. Obviously, the specific functional form of both $f(R)$ and $g(R)$ will depend on the equation of state $\omega$. For specific choices $\omega=0,1/3,1$ it can be shown, after some quick algebra, that the expressions (\ref{b1}), (\ref{b2}) and (\ref{b3}) reduce to those found in \cite{Bertolami} (modulo an overall factor two in the definition of $g(R)$).

\subsection{Power law $\Phi(\rho)$} \label{phi2}

Going beyond the above scenario, now we shall consider the case
\be
\Phi=a \rho^n,
\en
where $a$ is a constant and $n$ a parameter determining a family of Lagrangians. After some algebra one finds the functions characterizing the cosmological solutions of this model as
\bea
R&=&6 \dot{H}\left(1-\frac{3}{2} n(1+\omega) \right) \nonumber  \\
&+& 12H^2 \left(1+\frac{9}{4} n\omega(1+\omega) +\frac{9}{8} n(3n-2)(1+\omega)^2 \right) \\
f&=&2a\rho^n\Big[H^2 \left(3-\frac{15}{2}n+9n^2 +\frac{n}{2}(45n-9)\omega +\frac{27}{2} n^2 \omega^2 \right) \nonumber \\
&+& \frac{(1+3\omega)}{(1+\omega)} \dot{H} \left(1-\frac{3}{2}n(1+\omega)\right)\Big]  \\
g&=&\frac{2a\rho^{n-1}}{\kappa^2} \Big[\frac{H^2}{4}\left(-3n(3n+2) -9n^2 \omega \right) \nonumber \\
&-&\frac{1}{(1+\omega)} \dot{H} \left(1-\frac{3}{2} n(1+\omega)\right) \Big],
\ena
where the function $H$ has not been specified yet. For the particular case of quadratic cosmology, as defined by Eqs.(\ref{eq:H2a}) and (\ref{eq:Hda}) the formulae above read explicitly
\bea
R&=& \frac{\kappa^2 \rho}{4} \Big[(2-9n+27n^2) + 6(-1+9n^2)\omega +9n(1+3n)\omega^2 \nonumber \\
&+& \epsilon \frac{\rho}{\rho_0} \Big( (-4+27n^2)+6(-2+3n+9n^2)\omega \label{eq:Rrho} \\
&+&18n(1+\frac{3n}{2}) \omega^2 \Big) \Big] \nonumber \\
f&=&\frac{a\kappa^2 \rho^{n+1}}{4} \Big[ (2-7n+12n^2)+6(-1+n+5n^2)\omega \nonumber \\
&+&9n(1+2n) \omega^2 \nonumber \\
&+&\epsilon \frac{\rho}{\rho_0} \Big(4n(-1+3n)+3(-4+6n+10n^2)\omega \\
&+&18n(1+n)\omega^2 \Big) \Big] \nonumber \\
g&=& \frac{a\rho^n}{6} \Big[(6-15n-9n^2)-9n(1+n)\omega \nonumber \\
&+&\epsilon \frac{\rho}{\rho_n} \Big( (12-24n-9n^2)-9n(2+n)\omega \Big) \Big].
\ena
The equation (\ref{eq:Rrho}) for $R(\rho)$ is quadratic in $\rho$ and, therefore,  can be analytically inverted. Writing it formally as $R(\rho)=d\rho [b(n,\omega)+c(n,\omega)\rho]$ where $d$, $b(n,\omega)$, and $c(n,\omega)$ are constants following from (\ref{eq:Rrho}), one gets
\be \label{eq:rhoR}
\rho=\frac{b(n,\omega)}{2c(n,\omega)} \left[-1 \pm \sqrt{1+\frac{4c(n,\omega) R}{d b^2(n,\omega)}} \right],
\en
where the choice $\pm$ will depend on the particular model $\epsilon = \pm 1$ and equation of state $\omega$ chosen. To avoid troubles with the square root in (\ref{eq:rhoR}), from Eq.(\ref{eq:Rrho}) it follows that for $n \geq 1$ the constants present in the expression of $R(\rho)$ must satisfy $b>0$ and $\text{sign}(c)=\text{sign}(\epsilon)$ (assuming $\omega \geq 0$). Thus, if $\epsilon=+1$ ($c>0$), we have $R>0$ and
\be
\rho_{\epsilon=+1}=\frac{b(n,\omega)}{2c(n,\omega)} \left[-1 + \sqrt{1+\frac{4c(n,\omega) R}{d b^2(n,\omega)}} \right],
\en
and it is positive everywhere. On the other hand, if $\epsilon=-1$ ($c<0$) then one obtains
\be
\rho_{\epsilon=-1}=\frac{b(n,\omega)}{2c(n,\omega)} \left[-1 - \sqrt{1+\frac{4 c(n,\omega) R}{d b^2(n,\omega)}} \right].
\en
The term under the square root is always positive as the function $R(\rho)$ grows in the region $\rho \simeq 0$, attains a maximum at $\rho_M=-b/(2c)$ and changes from positive to negative at $\rho_m=-b/c$, where the term under the square root takes its minimum value (zero) before growing again.

Though explicitly solvable, the model $\Phi=a \rho^n$ has the curious property of not recovering a linear behavior for $f(R)$ in the $R\to 0$ limit. In fact, it generically behaves as
\be
\lim_{R\to 0} f(R) = R^{n+1} + O(R^{n+1})
\en
and therefore, only for $n=0$ the expected low-energy behavior is obtained. Though this makes it hard to justify the viability of this particular family of models, the point is that they are able to reproduce the background expansion history of quadratic cosmology.

\subsection{Power law $\Phi(\rho)$ with a constant} \label{phi3}

Let us consider a different type of $\Phi(\rho)$ relation combining the two previous proposals
\be \label{eq:Phin}
\Phi(\rho)=b+a \rho^n,
\en
where again, $a$ and $b$ are some constants and $n$ a parameter determining the family of models. For simplicity let us take the choice $n=1$. Here we focus directly on the particular case of quadratic cosmology. Following the same procedure as in the previous cases, we obtain the explicit expressions for the relevant functions to be
\bea
R(\rho)&=& \frac{2\kappa^2 \rho}{(b+a\rho)} \Big[ \Big(1+\epsilon \frac{\rho}{\rho_0} \Big) \Big((b+a\rho(1+\frac{9}{4}\omega(1+\omega))) \nonumber \\
&+&\frac{9}{8} (1+\omega)^2 \frac{a^2 \rho^2}{(b+a\rho)} \Big)  \label{eq:R3} \\
&-&\frac{3}{4}\Big(1+2\epsilon \frac{\rho}{\rho_0}\Big)(1+\omega) \Big(b+a\rho(1-\frac{3}{2}(1+\omega)) \Big)  \Big] \nonumber\\
f(\rho)&=& \frac{\kappa^2}{2} \Big[2\Big(1+\epsilon \frac{\rho}{\rho_0}\Big) \Big(b+\frac{9}{2} a\omega(1+\omega)\rho \nonumber \\
&+& \frac{3}{2}(1+\omega)\frac{a^2 \rho^2}{(b+a\rho)} \Big)  \label{eq:f3}\\
&-& \Big(1+2\epsilon \frac{\rho}{\rho_0} \Big) (1+3\omega) \Big(b+a\rho(1-\frac{3}{2}(1+\omega)) \Big) \Big] \nonumber \\
g(\rho)&=&\frac{1}{6} \Big[4\Big(1+\epsilon \frac{\rho}{\rho_0} \Big) \Big(\frac{9}{8}(1+\omega)\frac{a^2\rho^2}{(b+a\rho)} - a(3+\frac{9\omega}{4} ) \Big) \nonumber \\
&+&3 \Big(1+2\epsilon \frac{\rho}{\rho_0} \Big)(b +a\rho(1-\frac{3}{2}(1+\omega))) \Big] \label{eq:g3}
\ena
where the corresponding expressions in sections (\ref{phi1}) and (\ref{phi2}) are obtained when $a=0$ and $b=0$ (and $n=1$), respectively. A glance at equation (\ref{eq:R3}) for $R(\rho)$ confirms that it is not easy to obtain a simple closed expression for $\rho(R)$,  $f(R)$, and $g(R)$ for generic values of the parameters $a$, $b$ and $\omega$. The resolution, therefore must be done case-by-case. A particularly simple case is that of $\epsilon=+1$ and $\omega=0$. Taking for simplicity $a=b=\kappa^2=1$, and solving the equations (\ref{eq:R3}), (\ref{eq:f3}) and (\ref{eq:g3}) one finds that as $R\to 0$, $f(R)$ and $g(R)$ behave as
\bea
f(R)&=&R-\frac{4R^2}{\rho_0}+\frac{20 R^3}{\rho_0^2} + \ldots \\
g(R)&=& \frac{1}{2} - \frac{5R}{2\rho_0} + \frac{11R^2}{2\rho_0^2} + \ldots
\ena
where the factor $1/2$ appearing in the first term of $g(R)$ can be put to one simply by a redefinition of units. This shows that the right GR limit is recovered. Consequently, this scenario provides valuable models both in the full and relaxed regimes, being at the same time consistent with the right GR behavior for low curvatures. Models of the form (\ref{eq:Phin}) with $n \neq 1$ are also expected to be consistent.

To conclude we point out that more general constraints of the form

\be \label{eq:Phin}
\Phi(r)=b+a_1 \rho^{n_1} + a_2 \rho^{n_2} + \ldots
\en
with $b,a_1,a_2, \ldots$ some constants and $n_1, n_2, \ldots$ some parameters, can be analyzed in a similar way and are expected to modify, in the low-curvature regime, the coefficients multiplying the powers of the Ricci scalar in the corresponding expansion.

\section{Summary and conclusions}


In this work we have considered the problem of reconstructing any cosmological background history characterized by two given functions $H^2(\rho)$ and $\dot H(\rho)$ within the framework of Palatini $f(R)$ gravity theories with non-minimal curvature-matter coupling $g(R)$ and found that it can be solved in general. To obtain the solution, one must establish some relation between the functions $f(R)$ and $g(R)$, though this relation needs not be restricted to the case $\Phi \equiv f_R +2\kappa^2 g_R L_m=$ constant required in the metric approach \cite{Bertolami}.  If fact, since  the Palatini field equations are second-order, the choice $\Phi=\Phi(\rho)$ always leads to a solution. Since in the Palatini approach one avoids unnecessary a priori constraints on the $f(R)$ and $g(R)$ functions, this gives a greater freedom to study the dynamics of this kind of theories. The reconstruction program, therefore, appears to be more naturally implemented within the Palatini formulation than within the standard metric approach, which must be forced to behave like the Palatini one in order to find solutions.

We note that when the two functions $f(R)$ and $g(R)$ are given, then one obtains from (\ref{eq:trace}) a specific solution for $R(\rho)$, which determines the form of $\Phi(\rho)$ and of the pair ($H^2(\rho), \dot H(\rho)$). If, instead, one gives $H^2(\rho)$, which determines $\dot H(\rho)$, and $\Phi(\rho)$, then  $f(R)$, $g(R)$, and  $R(\rho)$ can be found. In both cases two inputs are necessary to completely define the cosmology. In the approach of \cite{Bertolami}, however, just one function $H^2(\rho)$ was necessary to obtain the functions $f(R)$ and $g(R)$, since the constant $\Phi_0$ can always be reabsorbed into a redefinition of the units used to measure $R$. This provides further support to the approach presented here over the constrained formulation of the metric case. 

We have also shown that the background expansion history of a particular Hubble function $H^2(\rho)$ can be reproduced using very different input functions $\Phi(\rho)$. However, the resulting $f(R)$ and $g(R)$ functions do not, in general, need to recover the expected behaviors $f(R)\sim R$ and $g(R)=$ constant as $R\to 0$. If this requirement is imposed, which is natural in order to have agreement with observations in non-cosmological scenarios, then the function  $\Phi(\rho)$ must be of the form $\Phi(\rho)\approx \Phi_0 + $ corrections, being $\Phi_0$ a constant. This provides a simple argument to constrain the freedom in the choice of  $\Phi(\rho)$ and confirms that the choice $\Phi(\rho)= \Phi_0 $  made in \cite{Bertolami} is a good one but not necessarily the only one.

\section*{Acknowledgments}

G.J.O. is supported by the Spanish grant FIS2011-29813-C02-02, the Consolider Program CPANPHY-1205388, the JAE-doc program of the Spanish Research Council (CSIC), and the i-LINK0780 grant of CSIC. D.R.-G. is supported by CNPq (Brazilian agency) through project No. 561069/2010-7 and acknowledges partial support from  project FIS2011-29813-C02-02. This work has also been supported by CNPq project No. 301137/2014-5. We are grateful to O. Bertolami and J. P\'aramos for useful comments.


\begin{thebibliography}{99}



\bibitem{Shiromizu:1999wj}
R. Maartens and K. Koyama,
Living Rev. Relativity \textbf{13}, 5 (2010);
  T.~Shiromizu, K.~-i.~Maeda, and M.~Sasaki,
  Phys.\ Rev.\ D {\bf 62}, 024012 (2000);
  E.~E.~Flanagan, S.~H.~H.~Tye, and I.~Wasserman,
  Phys.\ Rev.\ D {\bf 62}, 044039 (2000).

\bibitem{Singh06}
  A.~Ashtekar and P.~Singh,
  Class.\ Quant.\ Grav.\  {\bf 28}, 213001 (2011);
A. Ashtekar, A. Corichi, and P. Singh,
Phys. Rev. D \textbf{77}, 024046 (2008);
A. Ashtekar, T. Pawlowski, and P. Singh, 
Phys. Rev. Lett. \textbf{96}, 141301 (2006);
Phys. Rev. D \textbf{74}, 084003 (2006);
P. Singh, Phys. Rev. D \textbf{73}, 063508 (2006).

\bibitem{Bertolami}
O.~Bertolami and J.~P\'aramos,
  Phys.\ Rev.\ D {\bf 89}, 044012 (2014).

\bibitem{Olmo:2012yv}
  G.~J.~Olmo,
  {\it Introduction to Palatini theories of gravity and nonsingular cosmologies},
  Chapter of the book 'Open Questions in Cosmology', edited by Gonzalo J. Olmo (InTech Publishing, Rijeka, Croatia, 2012)
[arXiv:1212.6393].

\bibitem{Olmo:2008nf}
  G.~J.~Olmo and P.~Singh,
  JCAP {\bf 0901}, 030 (2009).

\bibitem{Borunda}
M. Borunda, B. Janssen, and M. Bastero-Gil, JCAP \textbf{0811}, 008 (2008);
Q. Exirifard and M. M. Sheikh-Jabbari, Phys. Lett. B \textbf{661}, 158 (2008).

\bibitem{Bertolami4}
O. Bertolami, C. G. Boehmer, T. Harko, and F. S. N. Lobo,
  Phys.\ Rev.\ D {\bf 75}, 104016 (2007);
O. Bertolami and J. P\'aramos,
Int. J. Geom. Meth. Mod. Phys. \textbf{11}, 1460003 (2014).

\bibitem{Olmo:2005hc}
  G.~J.~Olmo, Phys.\ Rev.\ Lett.\  {\bf 95}, 261102 (2005);
 Phys.\ Rev.\ D {\bf 72}, 083505 (2005).


\bibitem{Olmo:2011uz}
  G.~J.~Olmo,
  Int.\ J.\ Mod.\ Phys.\ D {\bf 20}, 413 (2011).

\bibitem{Olmo:2006zu}
  G.~J.~Olmo,
  Phys.\ Rev.\ Lett.\  {\bf 98}, 061101 (2007).

\bibitem{Olmo:2008ye}
  G.~J.~Olmo,
  Phys.\ Rev.\ D {\bf 77}, 084021 (2008).


\bibitem{MartinezAsencio:2012xn}
  J.~Martinez-Asencio, G.~J.~Olmo, and D.~Rubiera-Garcia,
  Phys.\ Rev.\ D {\bf 86}, 104010 (2012).

\bibitem{Lobo:2013vga}
  F.~S.~N.~Lobo, J.~Martinez-Asencio, G.~J.~Olmo, and D.~Rubiera-Garcia,
  Phys.\ Lett.\ B {\bf 731}, 163 (2014);
arXiv:1403.0105 [hep-th].

\bibitem{Olmo:2011sw}
  G.~J.~Olmo,
  JCAP {\bf 1110}, 018 (2011).

\bibitem{or}
G. J. Olmo and D. Rubiera-Garcia, Phys. Rev. D \textbf{86}, 044014 (2012);
Eur. Phys. J. C \textbf{72}, 2098 (2012);
Int. J. Mod. Phys. D \textbf{21}, 1250067 (2012);
JCAP \textbf{1402},  010 (2014);
F. S. N. Lobo, G. J. Olmo, and D. Rubiera-Garcia, JCAP \textbf{1307}, 011 (2013).

\bibitem{Lobo:2014fma}
  F.~S.~N.~Lobo, G.~J.~Olmo, and D.~Rubiera-Garcia,
Eur. Phys. J. C, in press [arXiv:1402.5099 [hep-th]].

\bibitem{Wheeler}
J. A. Wheeler, Phys. Rev. \textbf{97}, 511 (1955);
C. W. Misner and J. A. Wheeler,  Ann. Phys. \textbf{2}, 525 (1957).

\bibitem{Torsion}
G. J. Olmo and D. Rubiera-Garcia, Phys. Rev. D \textbf{88}, 084030 (2013) 084030.

\bibitem{Hawking}
S. W. Hawking and G. F. R. Ellis, \emph{The Large Scale Structure of Spacetime} (Cambridge University Press, Cambridge);
B. F. Schutz, Phys. Rev. D \textbf{2}, 2762 (1970);
J. D. Brown, Class. Quantum Gravity \textbf{10}, 1579 (1993).


\bibitem{Harko}
T.~Harko, T.~S.~Koivisto and F.~S.~N.~Lobo,
  Mod.\ Phys.\ Lett.\ A {\bf 26} (2011) 1467.


\bibitem{Bertolami2}
O. Bertolami, F. S. N. Lobo and J. P\'aramos, Phys. Rev. D \textbf{78}, 064036 (2008).


\bibitem{wald84}
R. M. Wald, {\it General Relativity} (University of Chicago Press, 1984).

\end{thebibliography}
\end{document}